\documentclass[useAMS,usenatbib]{mn2e}
\usepackage{graphicx}
\usepackage{amssymb}
\title[Global $m=1$ instabilities and lopsidedness in disc
galaxies]{Global $\textbf{m=1}$ instabilities and lopsidedness in disc
galaxies}
\author[V. Dury et al.]{V.~Dury$^1$\thanks{E-mail: 
Vanessa.Dury@UGent.be}, S.~De~Rijcke$^1$, Victor~P.~Debattista$^2$, H.
~Dejonghe$^1$\\$^1$Sterrenkundig Observatorium, Universiteit Gent, Krijgslaan 
281, S9, B-9000 Ghent, Belgium\\$^2$RCUK Fellow; Centre for 
Astrophysics, University of Central Lancashire, Preston, PR1 2HE, UK}
\begin{document}
\date{}
\pagerange{\pageref{firstpage}--\pageref{lastpage}} \pubyear{}
\maketitle
\label{firstpage}

\begin{abstract}
 Lopsidedness is common in spiral galaxies. Often, there is
no obvious external cause, such as an interaction with a nearby
galaxy, for such features. Alternatively, the lopsidedness may have an
internal cause, such as a dynamical instability. In order to explore
this idea, we have developed a computer code that searches for
self-consistent perturbations in razor-thin disc galaxies and
performed a thorough mode-analysis of a suite of dynamical models for
disc galaxies embedded in an inert dark-matter halo with varying
amounts of rotation and radial anisotropy.

Models with two equal-mass counter-rotating discs and fully rotating
models both show growing lopsided modes. For the counter-rotating
models, this is the well-known counter-rotating instability, becoming
weaker as the net rotation increases. The $m=1$ mode of the maximally
rotating models, on the other hand, becomes stronger with increasing
net rotation. This rotating $m=1$ mode is reminiscent of the
eccentricity instability in near-Keplerian discs.

To unravel the physical origin of these two different $m=1$
instabilities, we studied the individual stellar orbits in the
perturbed potential and found that the presence of the perturbation
gives rise to a very rich orbital behaviour. In the linear regime,
both instabilities are supported by aligned loop orbits. In the
non-linear regime, other orbit families exist that can help support
the modes. In terms of density waves, the counter-rotating $m=1$
mode is due to a purely growing Jeans-type instability. The rotating
$m=1$ mode, on the other hand, grows as a result of the swing
amplifier working inside the resonance cavity that extends from the
disc center out to the radius where non-rotating waves are stabilized
by the model's outwardly rising $Q$-profile.
\end{abstract}

\section{Introduction}
The stellar and/or gaseous discs of spiral galaxies are often affected
by large-scale asymmetries. About half of all late-type galaxies show
a lopsided structure that affects the whole disc \citep{rs94,
h98}. Based on near-infrared images of a sample of 149 disc galaxies,
\citet{b05} find that a large fraction of them have asymmetric stellar
discs. The strength of the lopsidedness does not correlate with the
presence of companions but, instead, correlates with the presence of
bars and spiral arms. They explore three different causes for the
lopsidedness: galaxy interactions, galaxy mergers, and gas
accretion. These authors favour the latter explanation, which indeed
can trigger strong lopsidedness if the gas in-fall is sufficiently
asymmetric. \citet{a06} analysed H{\sc i} surface density maps and
R-band images of 18 galaxies in the Eridanus group. All galaxies
showed significant lopsidedness in their H{\sc i} discs. Where the
stellar and gaseous discs overlap, their asymmetries are comparably
strong. Since the Eridanus group galaxies are more strongly lopsided
than field galaxies, these authors conclude that tidal interaction in
the group environment may contribute to generating lopsidedness in
disc galaxies. \citet{b1} studied kinematic lopsidedness in two spiral
galaxies and argued that it may be related to lopsidedness in the
potential. Alternatively, the stellar disc may lie off-centre in the
halo's gravitational potential well and spin in a sense retrograde to
its orbit about the halo centre \citep{ls98}. Lopsided structures are
not the prerogative of disc galaxies alone. In some nucleated dwarf
elliptical galaxies, for instance, the nucleus is displaced with
respect to the centre of the outer isophotes \citep{b3b}. While some
authors regard these displaced nuclei as being globular clusters
projected close to the galaxy photocenter \citep{acs}, in some cases,
such as the Fornax dwarf elliptical FCC046, there are clear
indications that the nucleus is an integral part of the galaxy's
stellar body and is displaced by a mechanism that affects the whole
galaxy \citep{b3}.

\citet{b4} used N-body simulations to investigate the stability of a
family of oblate elliptical galaxy models and found that a strong
lopsided instability occurs in models with small radial anisotropy and
strong counter-rotation. This instability may cause lopsidedness in
stellar systems with no or small net rotation, such as dwarf
ellipticals \citep{b3}. \citet{bc99} compiled a list of several types
of counter-rotation in galaxies of different morphological type. They
regard counter-rotation of stellar vs. stellar discs as the prevailing
type of counter-rotation since it is the end state of a galaxy with an
embedded counter-rotating star-forming gas disc. 

\citet{r92}
discovered in the S0 galaxy NGC4550 two distinct stellar components
rotating in opposite directions. Another example is the normal Sab
galaxy NGC7217 \citep{mk94}. \citet{v07} present the case of NGC5713,
a Sbc spiral galaxy in which 20~\% of the stars are on retrograde
orbits. Their data suggest that NGC5713 accreted neutral gas from its
surroundings on retrograde orbits. This gas was subsequently converted
into stars in a counter-rotating disc.  Less than 10~\% of all S0s
host counter-rotating stellar populations while counter-rotating gas
is found in roughly one quarter of them \citep{k96}. \citet{kf01} find
4 counter-rotators among a sample of 17 elliptical and lenticular
galaxies. They find no counter-rotation among 38 Sa-Sbc
galaxies. 

An axisymmetric galactic disc perturbed by a
constant lopsided halo potential causes a net lopsided distribution in
the disc, opposite to the perturbation halo potential
\citep{j97,j99}. \citet{ty00} suggest that counter-rotation can be
produced when a triaxial halo with an initially retrograde pattern
speed slowly changes to a pro-grade pattern speed. This is a situation
that probably does not occur very often. Moreover, this mechanism
requires evolution on long timescales of the order of $\sim
10^{10}$~yr. Overall, counter-rotation is detected only rarely in disc
galaxies.  Hence, the lopsidedness observed in so many disc galaxy is
most likely not caused by counter-rotation.

In this paper, we investigate the role played by instabilities in
generating lopsidedness in isolated disc galaxies using a
semi-analytic matrix method developed by \citet{b9}. More
specifically, we want to explore whether lopsided instabilities can be
triggered in fully rotating disc galaxies. In the next section, we
introduce the formalism underlying the computer code that we developed
to analyse the stability of a given dynamical model for a disc
galaxy. In section \ref{unperturb}, we present the unperturbed toy
galaxy models whose stability is analysed in section \ref{perturb}. We
investigate under what physical circumstances (i.e. degree of
counter-rotation and orbital anisotropy) lopsided structures can
spontaneously grow in these disc galaxy models. In section
\ref{physics}, we give a physical explanation for the self-consistent
growth of lopsided structures, based on the response of individual
stellar orbits to the growing instability. We summarise our
conclusions in section \ref{conclusions}.

\section{Searching for instabilities} \label{search}

We have developed a computer code to analyse the stability of
razor-thin stellar discs embedded in an axisymmetric or spherical dark
matter halo. The halo is assumed to be dynamically too hot to develop
any instabilities. This inert halo only enters the calculations by its
contribution to the global gravitational potential. We only consider
the stellar component of the disc and neglect the dynamical influence
of gas and dust.

We describe an instability as the superposition of a time-independent
axisymmetric equilibrium configuration and a perturbation that is
sufficiently small to warrant the linearisation of the Boltzmann
equation. The equilibrium configuration is characterised completely by
the global potential $V_0(r)$ and the distribution function
$f_0(E,J)$, with binding energy $E$ and angular momentum $J$. A
general perturbing potential can be expanded in a series of normal
modes of the form
\begin{equation}
V'(r,\theta , t)=V'(r)e^{i(m\theta -\omega t)}, \label{Vp}
\end{equation}
with a pattern speed $\Re(\omega)/m$ and a growth rate $\Im(\omega)$,
that, owing to the linearity of the relevant equations, can be studied
independently from each other. We write the response of the distribution
function to a perturbation as:
\begin{equation}
f(r,\theta,v_r,v_{\theta},t)=f_0(E,J)+f'(r,\theta,v_r,v_{\theta},t).
\end{equation}
The evolution of the perturbed part of the distribution function is
calculated using the linearised collision-less Boltzmann equation:
\begin{equation}\label{v1}
\frac{\partial f'}{\partial t}-[f',E]=[f_0,V'].
\end{equation}
We rewrite the right-hand side of the last equation as:
\begin{eqnarray}
[f_0,V,] & = &-\nabla_{\mathbf{v}} f_0 \cdot \nabla_{\mathbf{r}} V'
\nonumber \\ & =& -(\frac{\partial f_0}{\partial
E}\nabla_{\mathbf{v}}E+\frac{\partial f_0}{\partial
J}\nabla_{\mathbf{v}}J)\cdot\nabla_{\mathbf{r}}V' \nonumber \\ & = &
\frac{\partial f_0}{\partial
E}\mathbf{v}\cdot\nabla_{\mathbf{r}}V'-\frac{\partial f_0}{\partial
J}\frac{\partial V'}{\partial \theta}.
\end{eqnarray}
We also know that:
\begin{equation}
\nabla_{\mathbf{r}}V'\cdot\mathbf{v}=\frac{\rm{d} V'}{\rm{d}
t}-\frac{\partial V'}{\partial t}.
\end{equation}
With this last identity, equation (\ref{v1}) becomes:
\begin{equation}
\frac{\partial f'}{\partial t}-[f',E]=\frac{\partial f_0}{\partial
E}\frac{\rm{d} V'}{\rm{d} t}+i(\omega\frac{\partial f_0}{\partial
E}-m\frac{\partial f_0}{\partial J})V'. \label{dfpdt}
\end{equation}
The left-hand side of eq. (\ref{dfpdt}) is simply the total time
derivative of $f'$ along an unperturbed orbit. If we integrate
eq. (\ref{dfpdt}) along the unperturbed orbits, we immediately obtain
the response of the distribution function to the perturbing
potential given by eq. (\ref{Vp}):
\[f'(\mathbf{r}_0,\mathbf{v}_0;t_0)=\]
\begin{equation} \label{v2}
\frac{\partial f_0}{\partial
E}V'(\mathbf{r}_0;t_0)+i(\omega\frac{\partial f_0}{\partial
E}-m\frac{\partial f_0}{\partial J})\int_{-\infty}^{t_0}
V'(r)e^{i(m\theta-\omega t)}{\rm{d}}\,t.
\end{equation}
The integral in eq. (\ref{v2}) converges if the perturbation
disappears for $t \rightarrow -\infty$ and is growing sufficiently
fast in time ($\Im(\omega)>0$).

Along an unperturbed orbit, the radial coordinate $r$ is a periodic
function of time with angular frequency $\omega_r$, just like $v_r$ and
$v_{\theta}$. Because the mean value of $v_{\theta}$ can be different
from zero, $\theta$ will be the superposition of a periodic function
$\theta_p(t)$ and a uniform drift velocity $\omega_{\theta}$:
\begin{equation}
\theta=\omega_{\theta}t+\theta_p(t).
\end{equation}
We separate the part of the integrand in eq. (\ref{v2}) that is
periodic with frequency $\omega_r$ from the aperiodic part and expand
it in a Fourier series:
\begin{eqnarray} \label{four}
V'(r)e^{i(m\theta-\omega
t)}&=&I(t)e^{i(m\omega_{\theta}-\omega)t}\nonumber \\
&=&e^{i(m\omega_{\theta}-\omega)t}\sum_{l=-\infty}^{\infty}I_le^{il\omega_rt}.
\end{eqnarray}
The coefficients $I_l$ are given by
\begin{equation}
I_l = \frac{1}{T} \int_0^T I(t) e^{-il\omega_r t}\,{\rm d}t,
\end{equation}
where the integration extends over half a radial period, starting at
apocentre at $t=0$. 

The first term of the right-hand side of equation (\ref{v2}) is
rewritten as:
\begin{eqnarray}
\frac{\partial f_0}{\partial E}V'(\mathbf{r}_0;t_0)&=&\frac{\partial
f_0}{\partial E}e^{i(m\theta_0-\omega
t_0)}\int_{-\infty}^{0}{\rm{d}}[I(t)e^{i(m\omega_{\theta}-\omega)t}]
\nonumber \\ &=&\frac{\partial f_0}{\partial E}e^{i(m\theta_0-\omega
t_0)}\sum_{l=-\infty}^{\infty}I_l.
\end{eqnarray}

The orbits are integrated in the unperturbed
potential using a leapfrog method. Instead of using $E$ and $J$,
orbits in the unperturbed potential are catalogued by their apocentre
and pericentre distances, denoted by $r_+$ and $r_-$,
respectively. The sense of rotation is indicated by the sign of
$r_-$. The grid of the orbit catalogue is given by
$(r_+,r_-)\in[0,r_{\rm max}]\times[-r_+,r_+]$. We use a grid of
$100\times200$ cells. For every orbit, $\omega_r$ and
$\omega_{\theta}$ are determined, the Fourier expansion (\ref{four})
is performed up to the order of $l_{\rm max}=50$, and the coefficients are
stored. If we want to calculate the response of the distribution
function in the point $(\mathbf{r}_0,\mathbf{v}_0)$ in phase space at
time $t_0$, we choose an orbit from the catalogue with the correct
integrals of motion (i.e. $r_+$ and $r_-$) but passing through its
apocentre at $t=0$ so the actual orbit has an offset in time $t(r_0)$ and
azimuth $\theta_p(r_0)$ that must be taken into account. Therefore we also 
store a tabulation of $t(r)$ and $\theta_p(r)$.

 The response of the distribution 
function to the perturbation now assumes the following concise form:
\begin{eqnarray}
f'(\mathbf{r}_0,\mathbf{v}_0;t_0) &=& e^{i(m\theta_0-\omega t_0)}
\times \nonumber \\ && \hspace{-7em} \sum_{l=-\infty}^{\infty}
I_l\frac{(l\omega_r+m\omega_{\theta})\frac{\partial f_0}{\partial
E}-m\frac{\partial f_0}{\partial
J}}{l\omega_r+m\omega_{\theta}-\omega}e^{i(l\omega_rt(r_0)-m\theta_P(r_0))}.
\end{eqnarray}
Because we are searching for instabilities, and thus $\Im(\omega)>0$, we do 
not have to be concerned by the presence of resonances.
We can compute the perturbed density by integrating the perturbed
distribution function over the velocities up to the escape
velocity. For self-consistent perturbations, the gravitational
potential produced by this response density should equal the original
perturbing potential. Using the matrix method developed by \citet{b9},
the search for a self-consistent mode of order $m$ is reduced to an
eigenvalue problem. We adopt the same basis set of potential-density
couples as \citet{b9}. The response density generated by each of these
basic potentials can be expanded in terms of the basic density
distributions. This gives rise to a matrix $C(\omega)$ that contains
the coefficients of these expansions. Self-consistent perturbations
have the unique property that they have a pattern speed and growth
rate $\omega$ for which $C(\omega)$ has an eigenvalue $\lambda=1$. The
corresponding eigenvector contains the coefficients of the expansion
of the response density in terms of the base set of density
distributions. Thus, for a self-consistent perturbation of order $m$ we are 
left with the numerical search within the complex plane for a value of 
$\omega$ for which $C(\omega)$ has a unity eigenvalue, something that can be 
accomplished very efficiently using bisection. For a more detailed description 
of the method, we refer the reader to \citet{b9}.

\section{The unperturbed models} \label{unperturb}

\begin{figure}
\includegraphics[width=5cm, height=8.7cm, angle=270,origin=br]{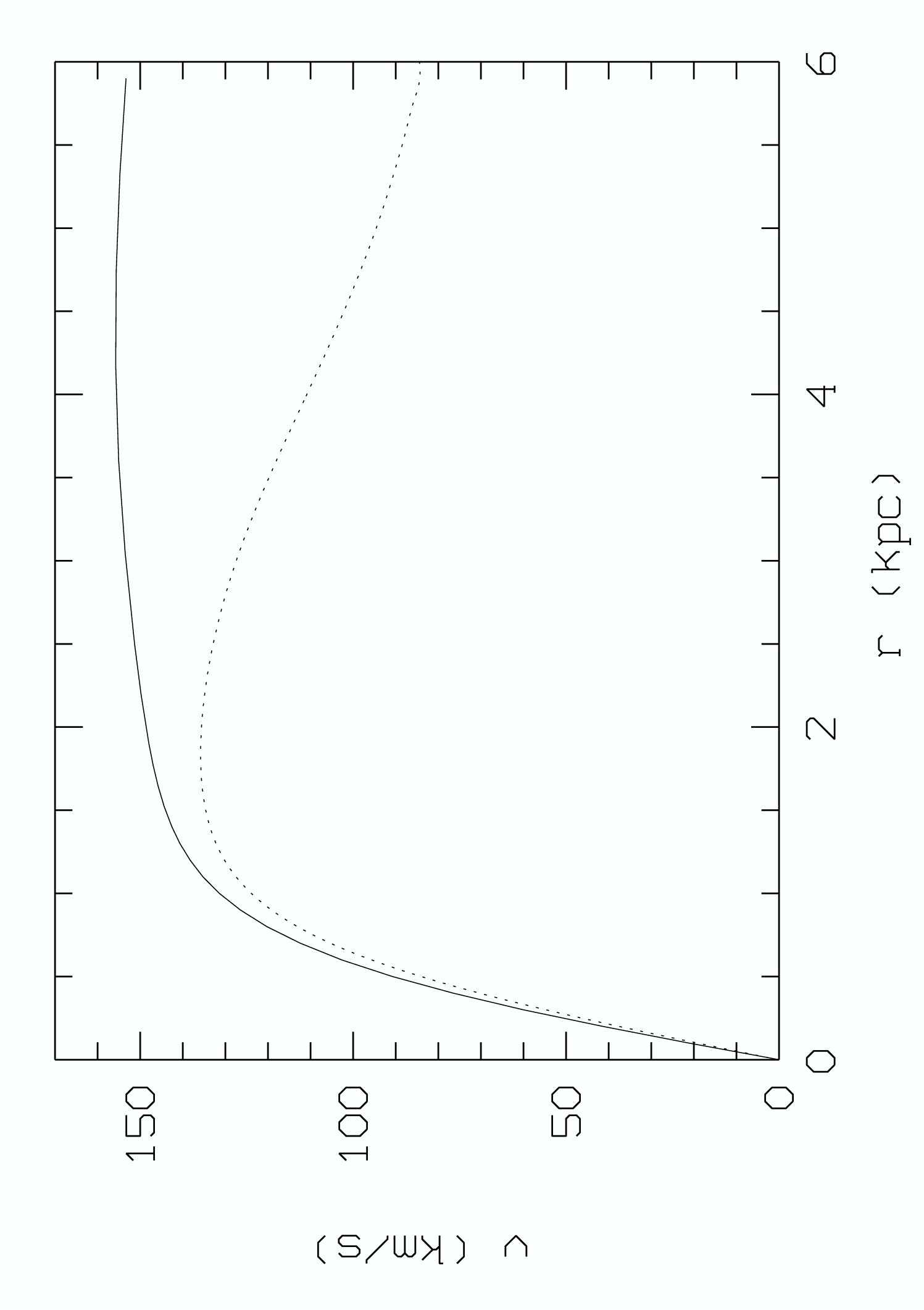}
\caption{Rotation curve of the unperturbed potential. The dotted line 
represents the contribution of the disc.}
\label{fig0}
\end{figure}

\begin{figure*}
\includegraphics[width=17cm]{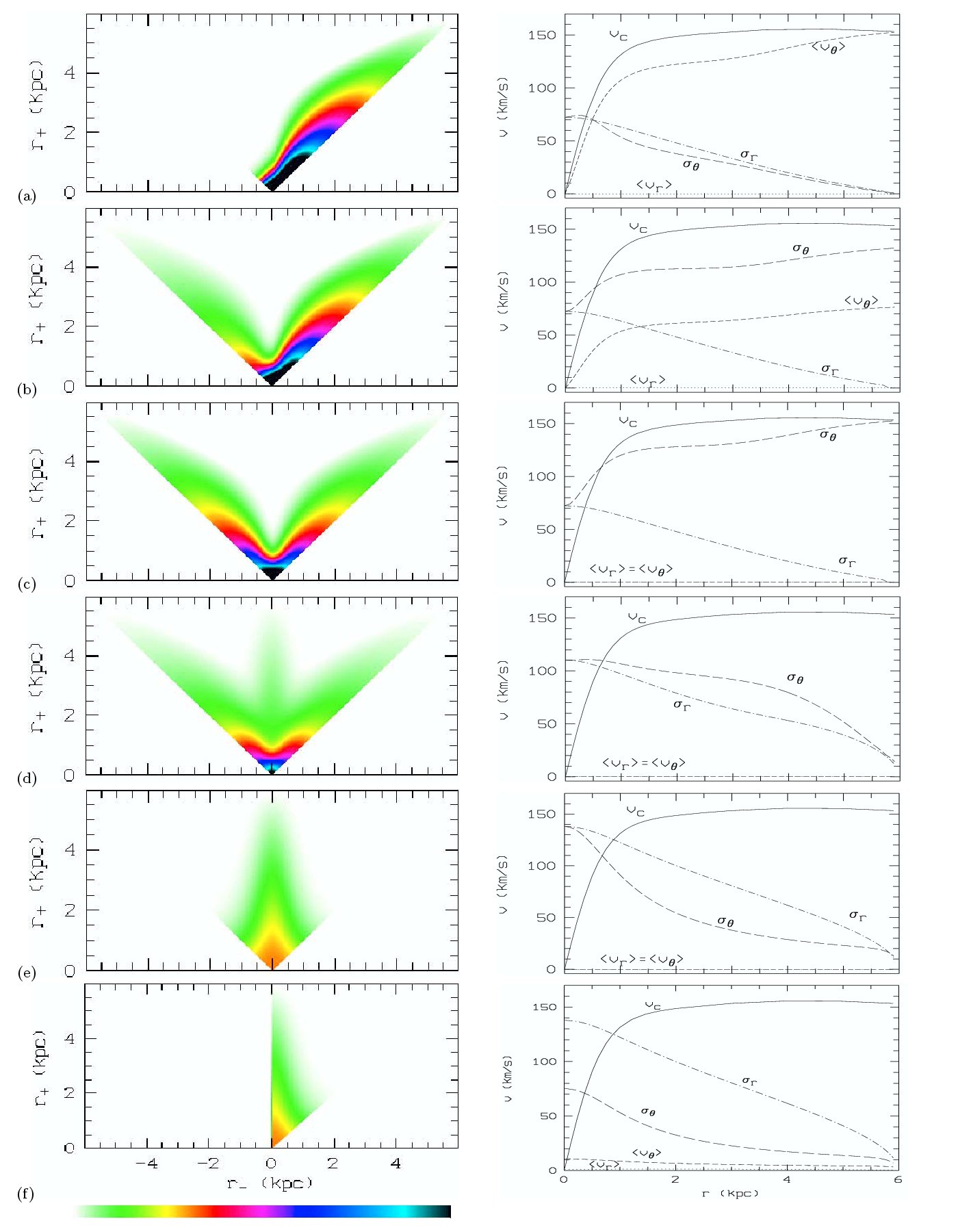}
\caption{Unperturbed distribution functions (left) and unperturbed
moments (right) of a few analysed models: (a) $f_0^{0,0}$, (b)
$f_0^{0.25,0}$, (c) $f_0^{0.5,0}$, (d) $f_0^{0.5,0.5}$, (e) $f_0^{0.5,1}$, 
(f) $f_0^{0,1}$. For the distribution functions the density is decreasing 
linear from black to white.}
\label{fig1}
\end{figure*}

A dynamical model for a disc galaxy embedded in a dark halo is
specified by the global potential of the system and the distribution
function of the stellar disc. For all the models that will be analysed
in this paper, we have chosenb> the same unperturbed potential and
unperturbed density but different distribution functions. The
unperturbed potential in the plane of the disc is given by:
\begin{equation}
V_0(r)=\frac{GM}{\sqrt{1+r^2}}+\frac{GM}{\sqrt{1+(\frac{r}{4.4})^2}},
\end{equation}
with $M=10^{10}M _{\odot}$ and $r$ expressed in kpc. This potential
produces a rotation curve that rises near the centre and becomes flat
further out in the disc, as can be seen in Fig.
\ref{fig0}. For the unperturbed density profile we choose an
approximately exponential profile:
\begin{equation}
\rho_0=\alpha e^{-1.3\sqrt{0.2+r^2}}. \label{rho0}
\end{equation}
The mass of the disc, and thus the mass of the halo, is determined by
$\alpha$. We impose an outer limit at $r_{\rm max}=6$ kpc. We have
chosen $\alpha$ so that the proportion $H/D$ of the total mass inside
the outer radius $r_{\rm max}$ for the halo and the disc is about
$2.5$. The disc is truncated by demanding that the distribution
function $f_0(E,J)$ is zero for orbits that venture outside this outer
limit. The distribution function is written as a linear combination of
a basic set of distribution functions. The coefficients of this
expansion are determined by a least-square fit to the density profile
given by eq. (\ref{rho0}). In all the models, the error on the fit to the 
mass density never exceeds $1\%$ of the central value. The contribution of 
the disc to the rotation curve is shown in Fig. \ref{fig0}.

We start with a strongly tangentially anisotropic model that is known
to develop strong spiral arms \citep{b9}. Its distribution function
$f_0^{0,0}(E,J)$ is shown in the left column of
Fig. \ref{fig1}a. Most stars in this model populate nearly circular
orbits that rotate in one direction. The unperturbed moments are shown
in the right column of Fig. \ref{fig1}a. Using the distribution
function $f_0^{1,0}(E,J)=f_0^{0,0}(E,-J)$, which has the same
unperturbed density as $f_0^{0,0}(E,J)$, one can easily construct
counter-rotating discs that all have the same density distribution. In
order to place a fraction $x$ of all stars on retrograde orbits, the
following distribution function can be used :
\begin{equation}
f_0^{x,0}(E,J)=(1-x)f_0^{0,0}(E,J)+xf_0^{1,0}(E,J).
\end{equation}
In Fig. \ref{fig1}b, we plotted the distribution function and
unperturbed moments of a model in which $25$ per cent of the stars
counter-rotate. In Fig. \ref{fig1}c, we present the properties of a
model consisting of two equally massive counter-rotating discs.

In the left column of Fig. \ref{fig1}e, we show the distribution
function $f_0^{0.5,1}(E,J)$ of a very radially anisotropic model that
is known to produce a strong bar instability \citep{b9}. Using this
model we can construct models with varying degrees of radial
anisotropy, as follows:
\begin{equation}
f_0^{0.5,y}(E,J)=(1-y)f_0^{0.5,0}(E,J)+yf_0^{0.5,1}(E,J). \label{f0rad}
\end{equation}
In Fig. \ref{fig1}d, we plotted the properties of a model consisting
of an even mix of $f_0^{0.5,0}$ and $f_0^{0.5,1}$ (i.e. $y=0.5$). We can also 
construct a similar family without counter-rotating discs:
\begin{equation}
f_0^{0,y}(E,J)=(1-y)f_0^{0,0}(E,J)+yf_0^{0,1}(E,J).
\end{equation}
In Fig. \ref{fig1}f, we plotted the distribution function of $f_0^{0,1}$.
 
\section{Lopsided instabilities} \label{perturb}

\subsection{The influence of counter-rotation}

\begin{figure*}
\includegraphics[width=17cm]{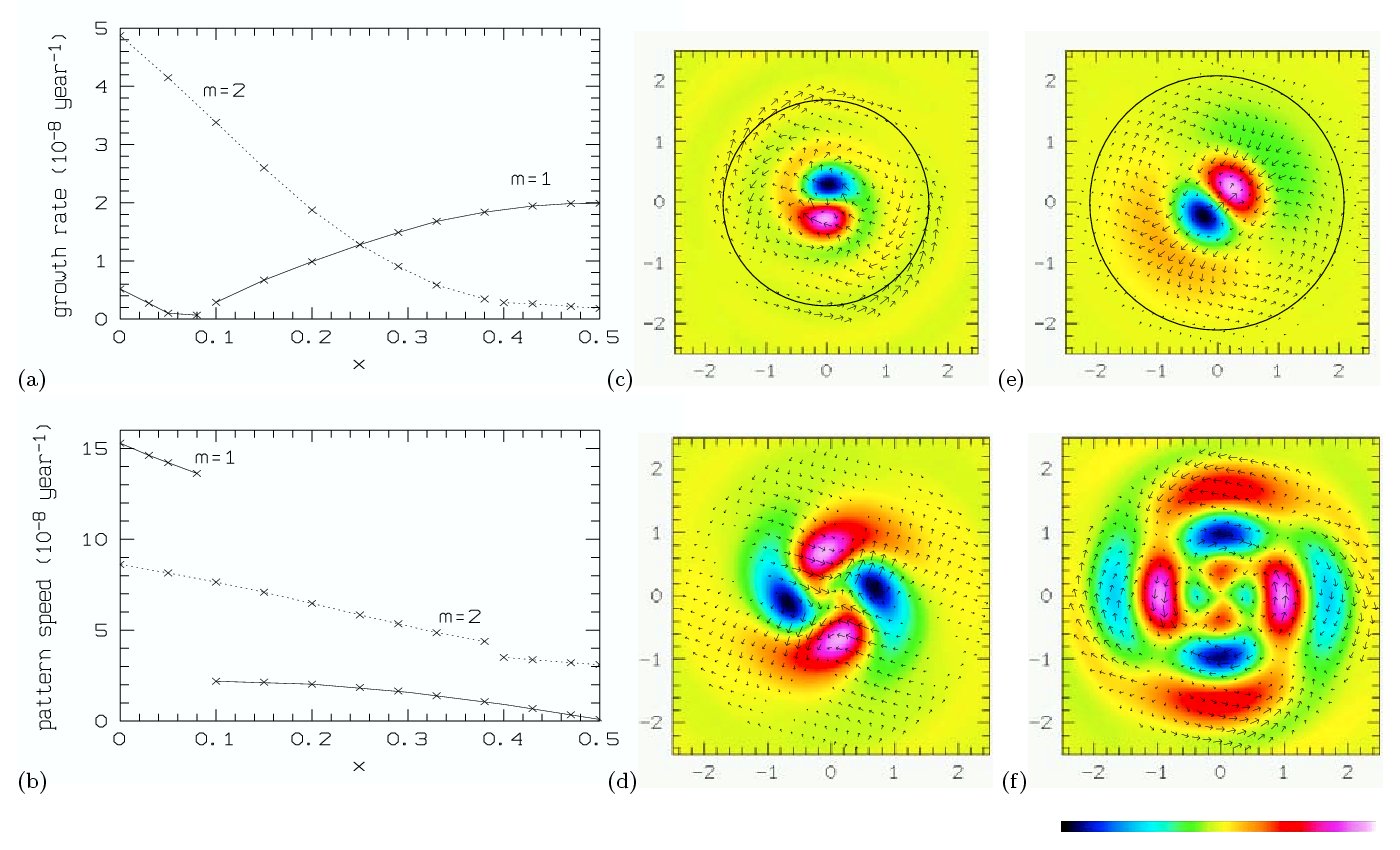}
\caption{Growth rate (a) and pattern speed (b) for the strongest $m=1$
(solid line) and $m=2$ (dotted line) instabilities as the fraction $x$
of counter-rotating stars changes. Discontinuities appear when the
nature of the instability changes. The perturbed density $\rho'
(\rho=\rho_0+\epsilon\rho')$ and velocity field $\mathbf{v}'
(\mathbf{v}=\mathbf{v}_0+\epsilon\mathbf{v}')$ is plotted out to a
radius of $2.5$ kpc: (c) rotating one-armed spiral for $f_0^{0,0}$ ,
(d) rotating two-armed spiral for $f_0^{0,0}$, (e) non-rotating
lopsided mode for $f_0^{0.5,0}$, (f) one of a pair of counter-rotating
bars for $f_0^{0.5,0}$. Over-densities are coloured white-red,
under-densities black-blue. In panels (c) and (e), we also indicate
the radius within which a growing $m=1$ mode is expected based on
density wave theory (see section \ref{physics} for a detailed
discussion).}
\label{fig2}
\end{figure*}

In Fig. \ref{fig2}, we show the growth rate and pattern speed of the
fastest growing $m=1$ and $m=2$ modes as the fraction $x$ of
counter-rotating stars for the models of the $f_0^{x,0}$ family
changes. Clearly, the fully counter-rotating $f_0^{0.5,0}$
model has a dominant non-rotating $m=1$ instability whereas the $m=2$
instability is virtually absent (actually, two mirror-image $m=2$
modes with the same growth rates and opposite pattern speeds occur in
the counter-rotating model). As the degree of counter-rotation
diminishes, the growth rate of the $m=1$ instability slowly declines
while it picks up a small pattern speed. At the same time, the $m=2$
instability increases in strength.  As long as more than one quarter
of all stars are on retrograde orbits, the model is dominated by a
slowly rotating lopsided mode.  Unexpectedly, around $x \approx 0.08$,
the nature of the $m=1$ mode changes. Another, this time rapidly
rotating, lopsided instability becomes the fastest growing $m=1$
mode. Its strength increases together with that of the $m=2$ mode as
the degree of counter-rotation vanishes.

The dominant $m=1$ mode we find in strongly counter-rotating models is
obviously the well-known counter-rotation instability. This
instability has been known since \citet{zh78} and has been studied
analytically \citep{s88,pp90,t05} and using N-body simulations
\citep{ms90,l90,b2,b4}. \citet{b4} do not detect it in systems rounder
than E6, which is mainly the result of their rounder models being
stabilised by a higher radial pressure. On the other hand,
\citet{ms90} find lopsidedness developing in systems as round as E1
but with negligible radial pressure. Partial rotation only introduces
a pattern speed in an otherwise purely growing instability. 

The counter-rotating bars we found, were also reported by \citet{b2},
\citet{l90} and \citet{f96}. \citet{b4} argued that they were the
result of non-linear orbit trapping in finite-amplitude spiral
disturbances. The fact that we found them shows that they are formed
through linear instabilities.

Recently, some authors found a lopsided instability in a normal
differentially rotating galactic disc \citep{sc07}. The lopsided
pattern precesses in the disc with a very slow pattern speed with no
preferred sense of precession.  The weaker $m=1$ mode that we found in
the rotating model has a certain sense of rotation and bears strong
resemblance to the so-called eccentricity instability that occurs in
gaseous and stellar near-Keplerian discs orbiting a central massive
object \citep{b6,star90,b7,ti98,l99,js01,b01,b5}. \citet{l99}
constructed models for disc galaxies with exponentially declining
surface density profiles embedded within a spherically symmetric dark
halo. These authors found the inner regions of such systems rapidly
develop a trailing one-armed spiral wave, even if the mass of the
central object is small. The first N-body example of a
rotating lopsided instability was found by \citet{s85} in a mass model
of our Galaxy without a halo component. \citet{e98} examined the
global stability of stellar power-law discs. They report a similar
rotating lopsided pattern in cut-out power-law discs, but found no
growing non-axisymmetric modes in the fully self-consistent power-law
discs. We provide the first theoretical evidence, based on a thorough
mode-analysis of a suite of self-consistent dynamical models
for disc galaxies embedded in a dark halo, that the eccentricity
instability can also occur in the fully prograde stellar discs of
spiral galaxies without an additional massive central component, such
as a compact bulge or super-massive black hole, and without
introducing an unresponsive central region in the disc.

\subsection{Anisotropy}

\begin{figure*}
\includegraphics[width=17cm]{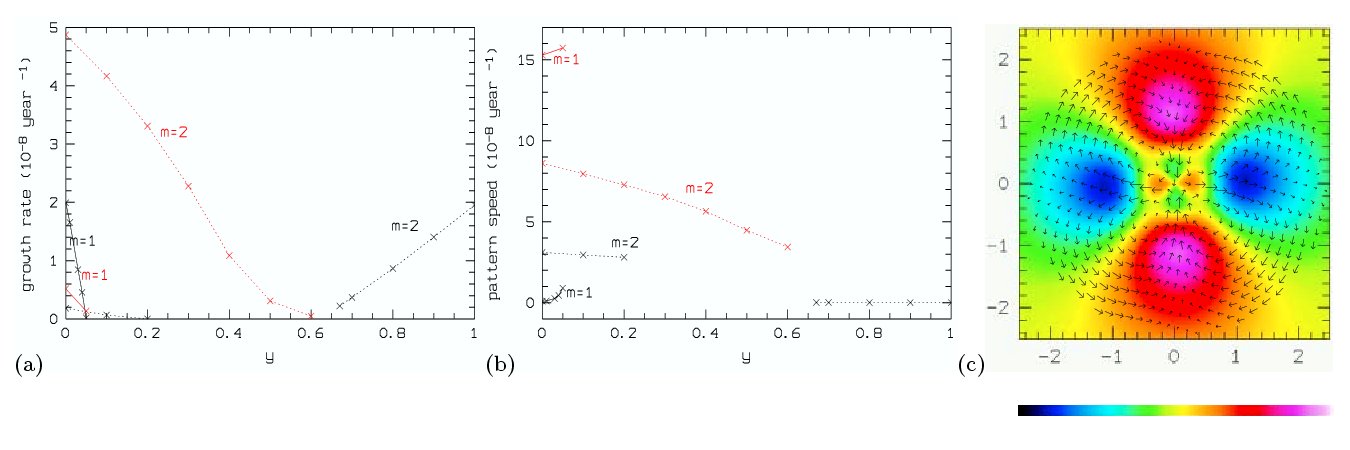}
\caption{Growth rate (a) and pattern speed (b) for
the strongest $m=1$ (solid line) and $m=2$ (dotted line) instabilities
as the fraction $y$ of stars on radial orbits changes for the
non-rotating (black) and rotating models (red). The perturbed density
$\rho' (\rho=\rho_0+\epsilon\rho')$ and velocity field $\mathbf{v}'
(\mathbf{v}=\mathbf{v}_0+\epsilon\mathbf{v}')$ is shown for model
$f_0^{0.5,1}$ in (c). Over-densities are coloured white-red,
under-densities black-blue. }
\label{fig4}
\end{figure*}

We now study the behaviour of the self-consistent $m=1$ and $m=2$ as the 
fraction $y$ of stars on radial orbits changes. The
variation of the growth rate and pattern speed is presented in Figs. 
\ref{fig4}a and \ref{fig4}b for the models $f_0^{0.5,y}$ (black) and 
$f_0^{0,y}$ (red). We start from model $f_0^{0.5,0}$ that is
known to develop a strong non-rotating $m=1$ instability and two twin
counter-rotating bar instabilities. As $y$, and thus the fraction of
stars on radial orbits, increases, the $m=1$ and $m=2$ instabilities
rapidly stabilise. The lopsided mode is the first to stabilise, at
$y\approx 0.05$. The model becomes fully stable against both $m=1$ and
$m=2$ modes at $y \approx 0.2$. For $y \gtrsim 0.6$, the model
develops a non-rotating bar instability that becomes stronger as
radial anisotropy increases. The perturbed density and velocity field
of the bar of the $f_0^{0.5,1}$ are shown in Fig. \ref{fig4}c. 
If we start from model $f_0^{0,0}$, the $m=1$ and $m=2$ instabilities also 
stabilise with increasing anisotropy. The $m=2$ instability, is the last one 
to stabilise and the model becomes stable for $y \gtrsim 0.6$.

From this exercise, it is clear that the mechanism that is responsible
for triggering the lopsided mode in the counter-rotating model relies
heavily on virtually all stars moving on near-circular orbits. Even a
relatively small contribution of stars on radial orbits makes it
impossible for the disc to develop the $m=1$ mode.

\subsection{Perturbed line-of-sight velocity fields}

\begin{figure*}
\includegraphics[width=17cm]{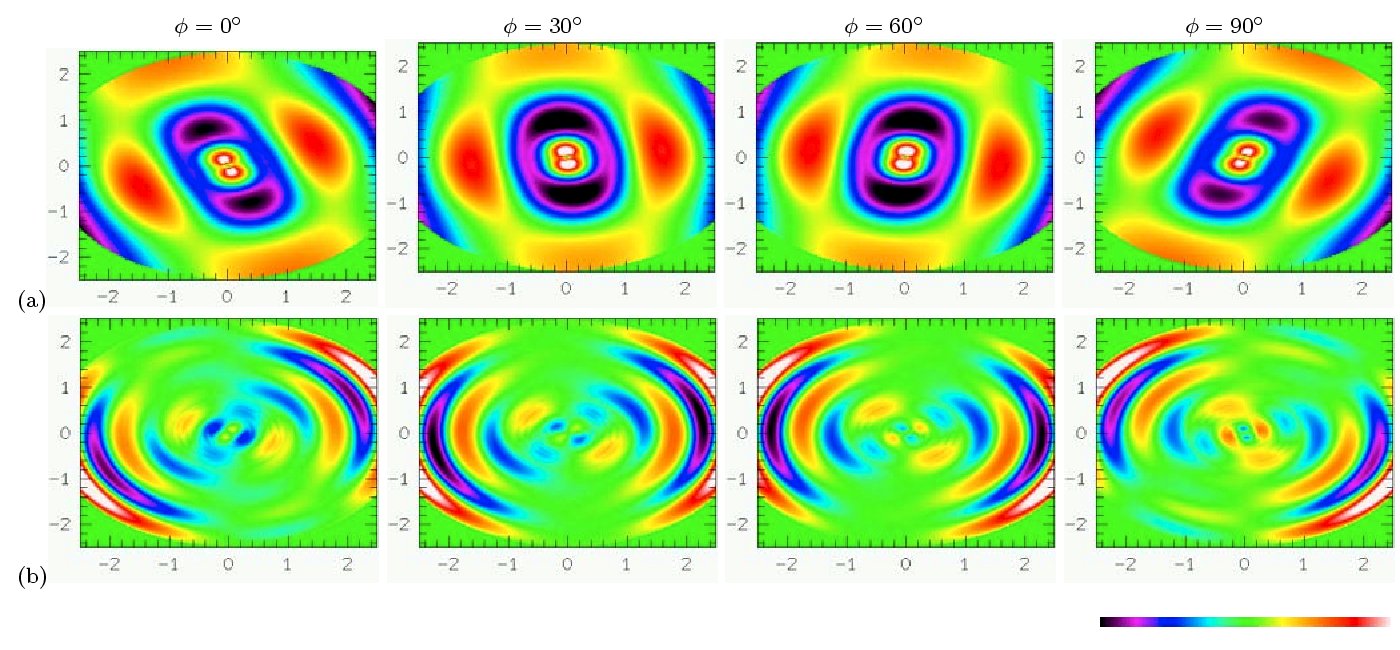}
\caption{Residual line-of-sight velocity field $v_{\rm los}'$ (with
$v_{\rm los}=v_{\rm los,0}+\epsilon v_{\rm los}')$ for the $m=1$
instabilities of (a) $f_0^{0.5,0}$ and (b) $f_0^{0,0}$ . All fields
have an inclination of 50$^\circ$ and a viewing angle $\phi$ varying
from 0$^\circ$ to 90$^\circ$ in steps of 30$^\circ$. Positive
velocities are coloured white-red, negative velocities black-blue.}
\label{fig3}
\end{figure*} 

To allow for a direct comparison of our models with observations, we
present in Fig. \ref{fig3} the perturbed line-of-sight velocity fields
for the extreme cases of exact counter-rotation (top row) and full
prograde rotation (bottom row).  The first-order perturbation to
the velocity $\mathbf{v}'$, corresponding to the perturbation $f'$ to
the distribution function $f_0$, is given by
\begin{equation}
\mathbf{v}'=\frac{\int\mathbf{v}f'{\rm
d}\mathbf{v}-\rho'\mathbf{v}_0}{\rho_0}
\end{equation}
This velocity perturbation is then projected onto the sky in order to
obtain the perturbation on the line-of-sight velocity, denoted by
$v_{\rm los}'$. The velocity fields in Fig. \ref{fig3} have an
inclination $i=50^\circ$ and a viewing angle $\phi$ ranging from
$0^\circ$ to $90^\circ$. The velocity fields are only plotted out to
a radius of 2.5~kpc since only inside this region is the density
perturbation noticeable. \citet{b10} showed that if the potential
contains a perturbation of harmonic number $m$ then the line-of-sight
velocity field contains $n=m-1$ and $n=m+1$ terms. Thus as expected,
we can see $n=0$ and $n=2$ components in the residual velocity fields
but there is no strong direct resemblance with the two galaxies
discussed by \citet{b1}, one of which shows a dominant $n=2$ velocity
perturbation while the other exhibits predominantly a $n=0$ structure.
 Of course, the residual velocity fields plotted in
Fig. \ref{fig3} are calculated using linear perturbation
theory. Instabilities in real galaxies are likely to be in the
non-linear regime. Therefore, this comparision is intended to be
indicative not definitive.

\section{Physical interpretation} \label{physics}

In order to unravel the physics behind the two distinct $m=1$ modes
found in section \ref{perturb}, we will study the orbits of stars that
move in the global perturbed potential:
\begin{equation}
V(r,\theta,t)=V_0(r)+\epsilon V'(r)e^{i(m\theta-\omega t)}. \label{testpert}
\end{equation}
In order to simplify the interpretation, we keep the amplitude of the
perturbation fixed, i.e. we set $\Im(\omega)=0$, and only consider its
pattern speed, $\Re(\omega)/m$. The prefactor $\epsilon$ is
determined by requiring that the maximum difference between the
perturbed and the unperturbed density nowhere exceeds 10~\% of the
unperturbed density. We then numerically evolve an ensemble of stars
in the perturbed potential using a leapfrog integrator. The goal is to
see how the orbits are affected by the perturbation and which
perturbed orbits help support the perturbation. For brevity, we will
henceforth refer to the $m=1$ instability of the $f_0^{0.5,0}$ model as the 
``non-rotating lopsided mode'', and to that of the $f_0^{0,0}$ model as the 
``rotating lopsided mode''.

\subsection{Non-rotating lopsided mode}

\begin{figure}
\includegraphics[width=8.5cm]{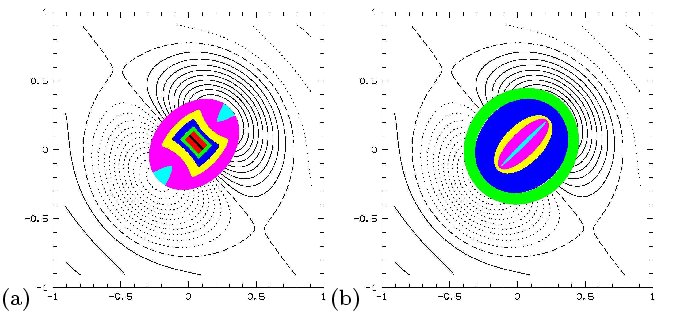}
\caption{Stellar orbits in the potential well of the non-rotating
lopsided mode. All orbits have the same energy but different angular momentum: 
Angular momentum increases in figure (a) from the black orbit ($J=0$) to the 
cyan orbit ($J=J_c$) and in figure (b) from the cyan orbit 
($J=J_c$) to the green orbit ($J=1.9~J_c$).}
\label{fig5}
\end{figure}

\begin{figure}
\includegraphics[width=8cm]{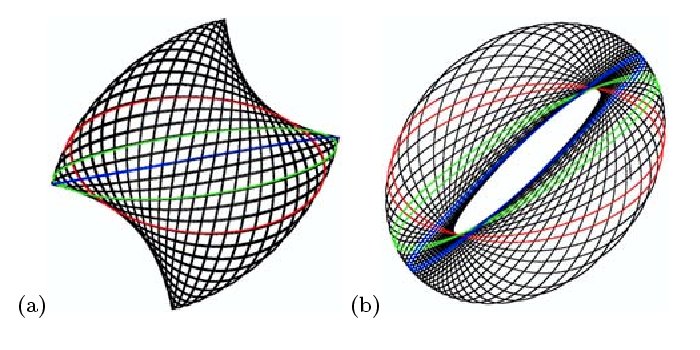}
\caption{(a) Butterfly orbit. (b) Loop orbit. To see the evolution in time, 
we indicated some parts of the orbit in colour. The first part red, the 
second part green and the last part blue.}
\label{fig6}
\end{figure}

In Fig. \ref{fig5}, we present some perturbed stellar orbits that
occupy the region in which the density perturbation is maximal. They
all share the same energy but are characterised by a different
angular momentum. We can easily recognise two general orbit
families~:~butterflies and loops. In Fig. \ref{fig6}a we show a more
detailed picture of a butterfly orbit. A butterfly can be viewed as a 
librating elliptical orbit with variable eccentricity. At the turning 
points of the libration, the orbit ellipticity becomes zero (this is 
evidenced by the red, green, and blue ellipses in Fig. \ref{fig6}a). As the 
angular momentum of the orbit is increased up to a critical point $J_c$, the 
two instances of zero ellipse orbit ellipticity coincide and the orbit fills 
an elliptical region. For still higher angular momentum, libration becomes 
rotation and the orbit becomes a loop, as can be seen in Fig. \ref{fig5}b \& 
Fig. \ref{fig6}b. Remarkably, both orbit families are also found by 
\citet{jal} for a disc galaxy model with a lopsided potential that is of 
St\"ackel form in elliptic coordinates and with two separate strong density 
cusps. They also found two other orbit families, nucleophilic bananas and 
horseshoe orbits. It is clear from their formulation that these two orbit 
families are associated with the cusps having diverging central densities 
which is why we do not find them in our models.

\subsubsection{Linear regime}

\begin{figure*}
\includegraphics[width=17cm]{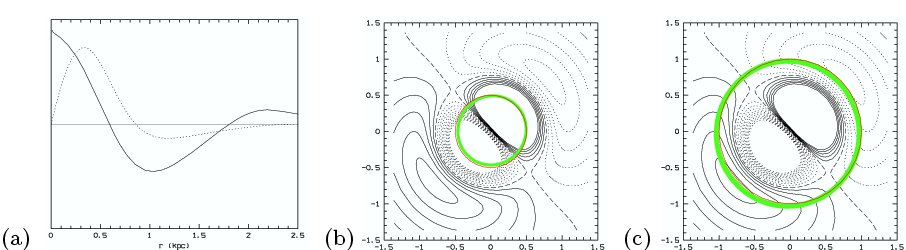}
\caption{(a) The sign of the quantity $C_2(r)$ (solid line) roughly
  traces that of the density perturbation (dotted line). Perturbed
  orbits (green) are displaced from the unperturbed orbit (red orbit):
  orbits with $C_2>0$ are shifted into the direction of the
  overdensity (b), those with $C_2<0$ are shifted into the opposite
  direction (c).}
\label{fig6e}
\end{figure*}

The mechanism that causes the non-rotating $m=1$ mode is now clear :
an infinitesimal $m=1$ perturbation will cause near-circular orbits to
become somewhat more elliptic and to shift towards the slight
overdensity, thus adding to this over-density, which in turn causes
other orbits to become more elongated and to shift, and so on. This
was already evident from section 3.3~3(a) of \citet{bt87} who used
perturbation theory in combination with the epicyclic approximation to
calculate the response of near-circular orbits to a general $m$-armed
perturbation.  Since this perturbation has no Lindblad resonances,
the condition that the sign of the quantity $C_2(r)=dV'(r)/dr + 2
V'(r)/r$ (represented by the red curve in Fig. \ref{fig6e}a) traces
that of the orbit displacement is fulfilled everywhere within the
stellar disc (see eq. (3-120b) of \citet{bt87}) and the perturbed
near-circular orbits will all strengthen the $m=1$ mode. Tangential
orbits with $C_2>0$ are shifted into the direction of the overdensity
(Fig. \ref{fig6e}b), those with $C_2<0$ are shifted into the opposite
direction (Fig. \ref{fig6e}c). As a consequence, the sign $C_2$ also
roughly follows that of the density perturbation. 

With a maximum density contrast of 10\%, the orbits plotted in
Figs. \ref{fig5} and \ref{fig6} are not per se in the linear
regime. This was done, since we here only wish to illustrate the
mechanism that induces the instability to grow, for clarity~:~a truly
infinitesimal perturbation would result in an equally infinitesimal
and therefore nearly invisible shift of the tangential orbits.

\begin{figure}
\includegraphics[width=8cm]{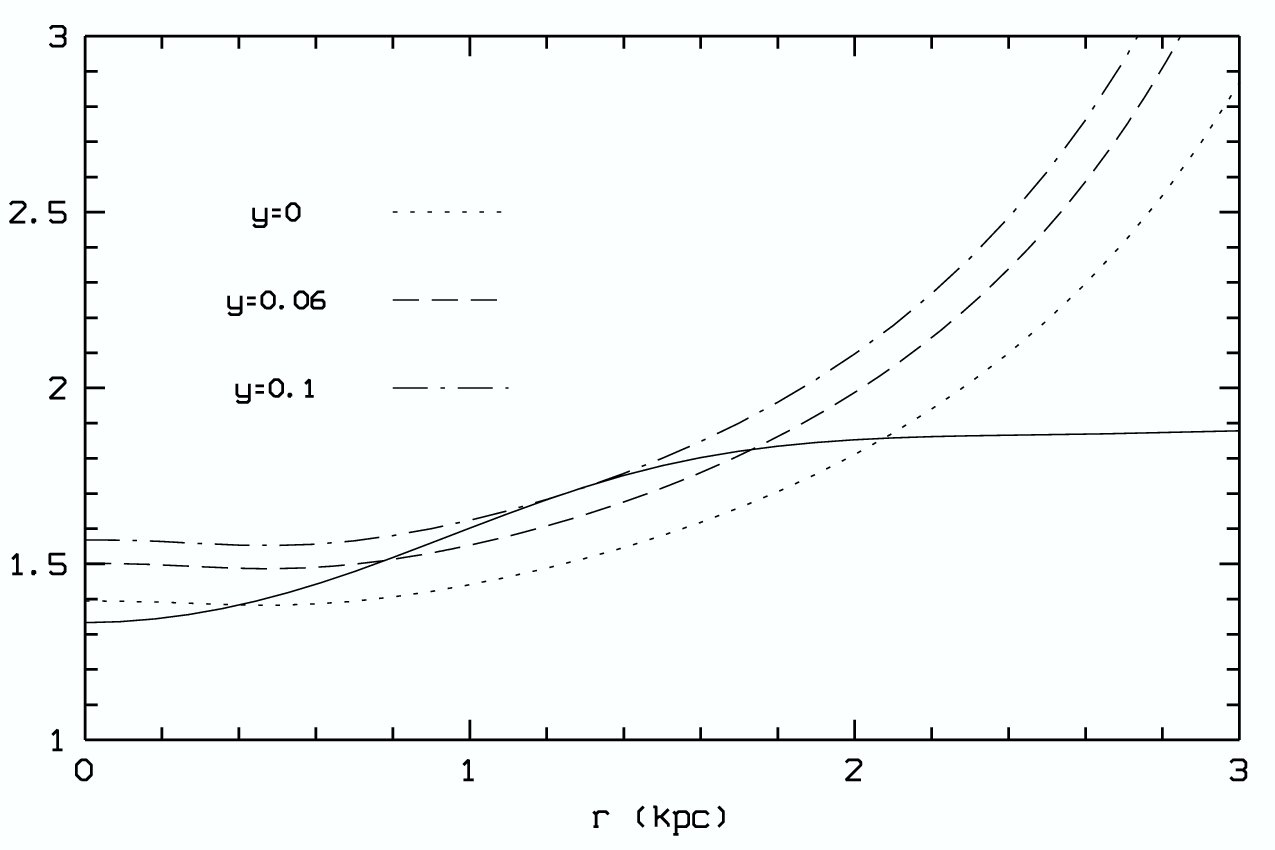}
\caption{In a counter-rotating disc, purely growing WKB waves may
develop when $Q(r)$ (dashed line) $\le
\frac{\kappa(r)^2}{\kappa(r)^2-\Omega(r)^2}$ (solid line). The
quantity $\frac{\kappa(r)^2}{\kappa(r)^2-\Omega(r)^2}$ is plotted
using a full line. The $Q-$ profiles of several counter-rotating
models, with increasing radial anisotropy as quantified by the
parameter $y$ in eq. (\ref{f0rad}), are overplotted (see legend in the
figure). Models with $y \gtrsim 0.06$ are stable, both according to
Palmer's stability criterion and our mode analysis.  }
\label{figQ}
\end{figure}

It is instructive to interpret this instability not only on the level
of stellar orbits but also in terms of density waves. As shown in
Palmer's book on dynamical instabilities \citep{p94}, a razor-thin
disc consisting of two equal-mass counter-rotating stellar populations
may develop purely growing one-armed ($m=1$) WKB waves if the Toomre
parameter, $Q=\frac{\sigma_r\kappa}{3.36G\rho_0}$,
fails to satisfy the local stability criterion
\begin{equation}
Q(r) \ge \frac{\kappa(r)^2}{\kappa(r)^2-\Omega(r)^2}.
\end{equation}
This is very similar to the well-known criterion $Q \ge 1$ for local
stability against purely growing $m=0$ waves. The precise form of this
stability criterion depends on three approximations, {\em (i)} the WKB
approximation for tightly wound spiral waves, {\em (ii)} the epicycle
description for near-circular stellar orbits, and {\em (iii)} a
Gaussian distribution function. None of these necessarily applies to
the models presented in this paper. Nonetheless, we used the $Q(r)$,
$\kappa(r)$, and $\Omega(r)$ profiles of the fully counter-rotating
$f_0^{0.5,y}$ models to evaluate this criterion as a function of
radius, where $Q$ depends on the parameter $y$ via the radial
velocity dispersion (Fig. \ref{figQ}). We only check the fully
counter-rotating models since only in this case can an analytical
stability criterion be derived. It is clear from Figs. \ref{fig2}(e)
\& \ref{figQ} that the region where, according to the local stability
criterion, purely growing waves may develop roughly coincides with the
region in which the non-rotating lopsided mode resides (i.e. inside a
radius $r \lesssim 2.1$~kpc).

As a further test, we checked whether the model that, according to our
mode-analysis, is the first to be stabilized by increasing radial
anistropy, is also stable according to Palmer's criterion. This is
done by increasing the parameter $y$ in eq. (\ref{f0rad}). It is clear
from Fig. \ref{figQ} that the $Q(r)$ profile of the first stable model
according to our mode-analysis, the one with $y \approx 0.06$ (see
fig. \ref{fig4}a), is also very close to the line of stability
according to Palmer's criterion. For $y \ge 0.1$, the models are
definitely stable both according to Palmer's criterion and our
mode-analysis. Note that Palmer's stability criterion is a sufficient
one~:~satisfying it implies stability, not satisfying it does not
necessarily imply instability. Given the reasonable agreement between
Palmer's analysis and our mode analysis concerning the line of
stability and the spatial extent of the $m=1$ pattern, we are led to
the conclusion that the purely growing lopsided mode we find in
counter-rotating discs is caused by a local Jeans-type instability.

\subsubsection{Non-linear regime}

 Figs. \ref{fig5} and \ref{fig6} reveal a feature that is not
captured by our linear mode-analysis. Once the perturbation is strong
enough, radial orbits are deformed into the new family of butterfly
orbits whose centres of gravity are also shifted in the direction of
the over-density. Thus, this orbit family may potentially contribute
to the $m=1$ perturbation but only after the amplitude of the
perturbation has become large enough, see also \citet{jal}. For an
infinitesimal perturbation, radial orbits do not contribute to the
growth of the instability (their presence is even detrimental to the
instability's growth in the linear regime, see below).  

\begin{figure}
\includegraphics[width=8.5cm]{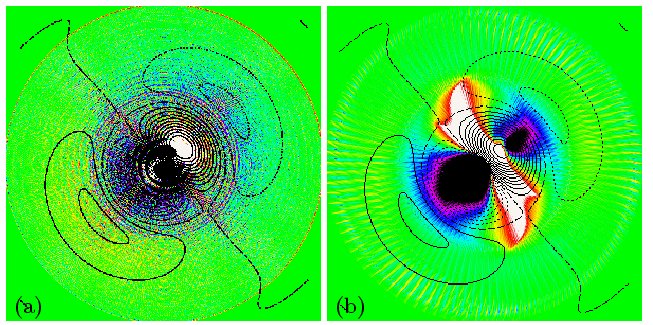}
\caption{The change in the density distribution of an ensemble of
orbits, $\delta\!\rho$, as defined by eq. (\ref{dro}), due to the
non-rotating $m=1$ mode. Panel (a) : an ensemble of circular
unperturbed orbits; panel (b) : an ensemble of radial unperturbed
orbits. Negative values of $\delta\!\rho$ are coloured black/blue;
positive values in white/yellow. The contours of the perturbed density
of the $m=1$ mode are also plotted (solid line: over-density, dotted
line: under-density). See text for a discussion of this figure.}
\label{fig7}
\end{figure}

Further evidence for this mechanism can be gleaned from the following
exercise. A test particle  with initial conditions
$(\mathbf{r}_0,\mathbf{v}_0)$ on an orbit in the unperturbed system
 with potential $V_0(r)$ over time fills a certain area, which,
in general will have the form of an annulus (with circular and
straight-line orbits as extremes). For the same initial conditions
 $(\mathbf{r}_0,\mathbf{v}_0)$, the test particle's orbit in the
perturbed potential  $V(r)$  will fill a differently shaped
area. The difference between the density distribution of the perturbed
orbit and that of the unperturbed orbit gives an idea of how the
density of a galaxy made up of an ensemble of unperturbed orbits will
change under the influence of the given perturbation. In
Fig. \ref{fig7}, we show how the density distribution of an
ensemble of circular orbits (\ref{fig7}a) or radial orbits
(\ref{fig7}b) changes due to the perturbation given by
eq. (\ref{testpert}); i.e., we plot the quantity
\begin{equation} \delta\!\rho(r,\theta) = \sum_{\rm orbit}\left(
\rho_{\rm orbit, pert}(r,\theta) - \rho_{\rm orbit,
unpert}(r,\theta)\right) \label{dro}
\end{equation}
where the sum runs over an ensemble of circular unperturbed orbits
with different radii and with the phases of the starting points of the
orbit integrations distributed uniformly over the interval [0,2$\pi$[
(circles in Fig. \ref{fig7}a and spokes in Fig. \ref{fig7}b are caused
by the finite number of orbits). Negative values of $\delta\!\rho$ are
coloured black/blue; positive values in white/yellow. Clearly, the
regions where $\delta\!\rho>0$ coincide with the over-densities of the
$m=1$ mode (full line contours) whereas the regions where
$\delta\!\rho<0$ coincide with the mode's under-densities (dotted line
contours). It is obvious from Fig. \ref{fig7}b that the radial orbits
are not nearly as cooperative  despite the fact that they are
slightly displaced towards the inner overdensity. As an ensemble,
they do not react to the imposed perturbation in a way that would tend
to strengthen it  since their long axes are oriented
perpendicularly to the inner lopsidedness leading rather to a $m=2$
feature. The (near-)circular orbits are clearly the backbone of this
$m=1$ mode and, as we know from Fig. \ref{fig4}a, even a smidgen of
stars on radial orbits is enough to stabilise the system against this
instability.

\subsection{Rotating lopsided mode}

\begin{figure*}
\includegraphics[width=17cm]{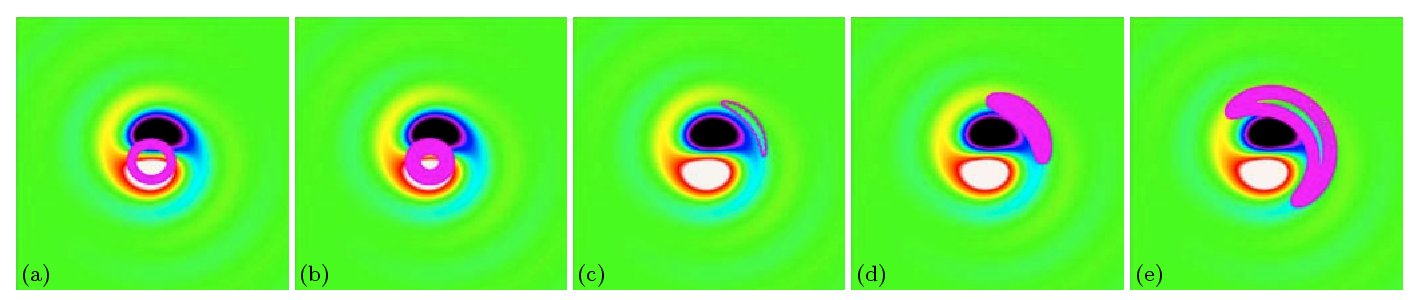}
\caption{Stellar orbits in the force field of the rotating lopsided
mode, as seen in a reference frame rotating with the same pattern
speed as the lopsided mode. We find two orbit families : (a) \& (b)
loop orbits, and (c), (d) \& (e) banana orbits.}
\label{fig8}
\end{figure*}

In Fig. \ref{fig8}, we present five perturbed stellar orbits that
occupy the region in which the density perturbation due to the
rotating lopsided mode is maximal. The orbits are plotted in a
reference frame that rotates with the pattern speed of the $m=1$
mode. Again we observe loop orbits that are displaced into the
direction of the over-density and that can support the lopsided
structure (Fig. \ref{fig8}a \& \ref{fig8}b). At larger radii, the
stellar orbits attain a banana shape  in this reference
frame. These orbits occupy a region outside the mode's main
under-density and seem to be connected to the one-armed spiral
(Fig. \ref{fig8}c, \ref{fig8}d \& \ref{fig8}e). Unlike in the previous
case, where aligned loop orbits were almost solely responsible for
creating the instability, here the two orbit families fulfil different
tasks. The aligned loop orbits support the inner lobes of the lopsided
structure whereas the banana orbits make up the outer one-armed
spiral.

\subsubsection{Linear regime}

\begin{figure*}
\includegraphics[width=17cm]{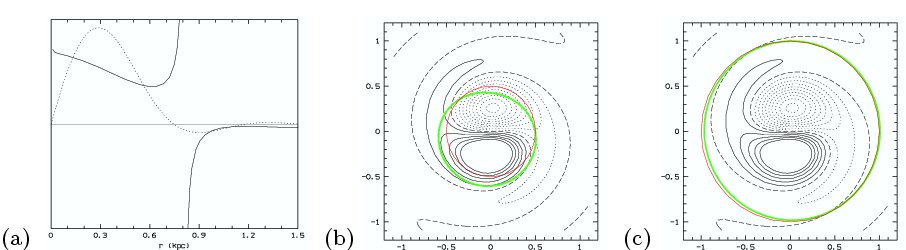}
\caption{(a) The sign of the quantity $C_2(r)$ (solid line) roughly
  traces that of the density perturbation (dotted line). Perturbed
  orbits (green) are displaced from the unperturbed orbit (red
  orbit). Orbits with $C_2>0$ are shifted into the direction of the
  overdensity (b) and those with $C_2<0$ are shifted into the opposite
  direction (c).}
\label{fig8b}
\end{figure*}

 For the lopsided mode with pattern speed $\Omega_p$, the quantity
$C_2(r)$ from \citet{bt87} becomes:
\begin{equation}
C_2(r)=\frac{1}{\kappa^2-m^2(\Omega-\Omega_p)^2}\left(\frac{{\rm
d}V'(r)}{{\rm d}r} + 2\frac{\Omega V'(r)}{r(\Omega-\Omega_p)}\right).
\end{equation} 
The sign of $C_2$ again traces that of the orbit displacement and,
with less fidelity, of the density perturbation
(Fig. \ref{fig8b}a). Orbits inside the corotation radius are shifted
into the direction of the overdensity $(C_2>0)$
(Fig. \ref{fig8b}b). At corotation, $C_2$ diverges. Orbits near
corotation become banana orbits. The corotation radius of the mode is
at $r_c=0.8$ kpc, which coincides with the position of the one-armed
spiral. Outside corotation, the sign of $C_2$ changes and orbits are
shifted into the other direction (Fig. \ref{fig8b}c).

The interpretation of this instability in terms of waves is somewhat
subtle. In the case of a bar instability, the radial extent of the
$m=2$ pattern is determined by  the largest radius out to which
the most slowly rotating wave that avoids having an inner Lindblad
resonance (ILR) can travel. This constraint sets the size of the
resonance cavity within which the pattern can grow by swing
amplification. However, one-armed waves do not have an ILR,
irrespective of their pattern speed. The radial extent of the $m=1$
pattern is set by the largest radius out to which non-rotating wave
packets can travel. This radius is determined by the disc's dispersion
relation. We used the WKB dispersion relation (eq. (6-46) of
\citet{bt87} or eq. (12.82) of \citet{p94}) together with the $Q(r)$,
$\kappa(r)$, and $\Omega(r)$ profiles of the $f_0^{0,0}$ model to
estimate the region in which non-rotating one-armed waves are allowed
to exist. The $Q(r)$ profile of the fully rotating $f_0^{0,0}$ model
rises outwardly, limiting the extent of non-rotating waves to some
finite radius, well within the disc. In the case of the $f_0^{0,0}$
model, for a zero pattern speed, the dispersion relation has two
branches (the long-wave and the short-wave branch) of incoming and
outgoing leading and trailing waves for radii smaller than
approximately 1.7~kpc. This sets the dimension of the resonance cavity
within which the pattern can grow through swing amplification. All
$m=1$ waves can propagate into the galaxy center where incoming
trailing wave packets are reflected as outgoing leading wave packets,
closing the feedback loop. Moreover, a more steeply rising rotation
curve will suppress the $m=2$ mode while leaving the $m=1$ mode, which
has no inner Lindblad resonance, largely intact.

Thus, following \citet{e98} who already proposed this mechanism as the
cause of the $m=1$ modes found in cut-out power-law discs, we propose
swing amplification as the physical cause of the one-armed mode in
this rotating model.  

\subsubsection{Non-linear regime}

\begin{figure}
\includegraphics[width=8.5cm]{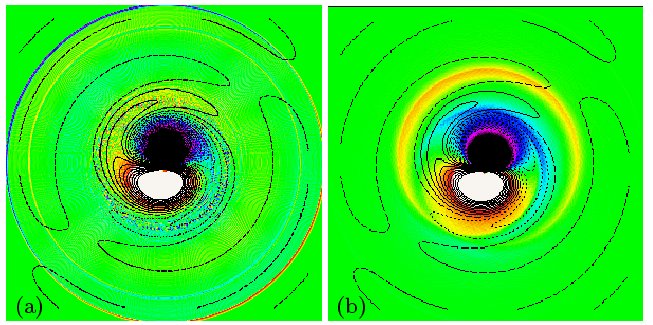}
\caption{The change in the density distribution due to the
rotating $m=1$ mode. Panel (a) : an ensemble of circular
unperturbed orbits; panel (b) : an ensemble of radial unperturbed
orbits. Negative values are coloured black/blue;
positive values in white/yellow. The contours of the perturbed density
of the $m=1$ mode are also plotted (solid line: over-density, dotted
line: under-density)}
\label{fig9}[width=17cm]
\end{figure}

In Fig. \ref{fig9}, we show how the density distribution, i.e. the
quantity $\delta\!\rho$, defined by eq. (\ref{dro}), of an ensemble of
circular (panel (a)) and of radial (panel (b)) orbits changes due to
the rotating $m=1$ mode. We now see that the circular orbits and the
radial orbits both support the lopsided structure. This is because
radial orbits as well as the inner circular orbits become loop orbits
that are aligned with the lopsided mode. If we change the fraction of 
stars on radial orbits, we have a slower stabilisation of the $m=1$ mode 
than in the non-rotating models (see Fig. \ref{fig4}a). 

\begin{figure}
\hspace*{7em}
\includegraphics[width=5cm]{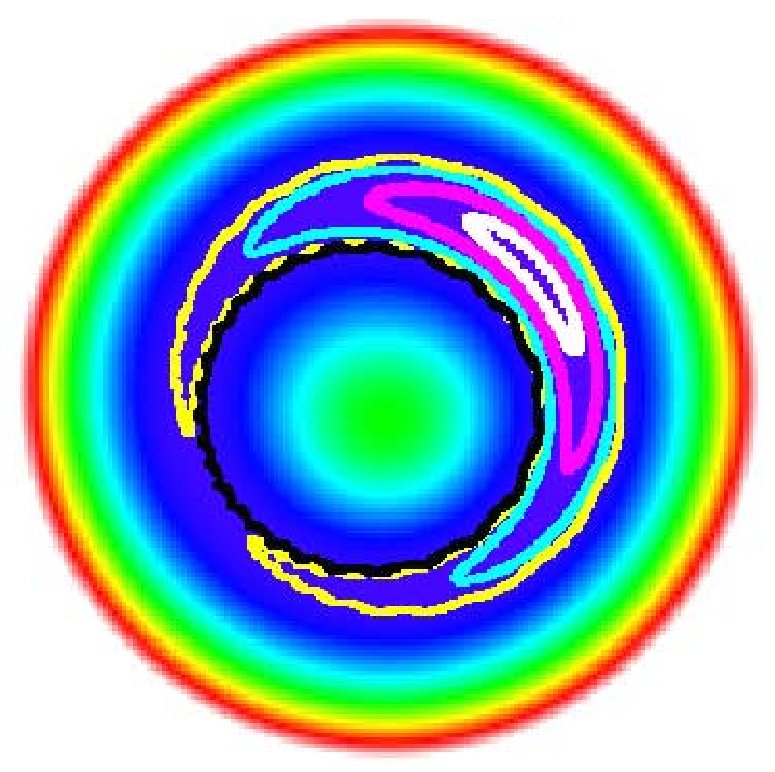}
\caption{A few banana orbits in a frame rotating with the lopsided mode. All 
orbits start from co-rotation with the same energy and angular momentum, but 
at a different angle $\theta$: white ($\theta=45^\circ$), purple 
($\theta=90^\circ$), cyan ($\theta=135^\circ$), yellow 
($\theta=180^\circ$), black ($\theta=225^\circ$). The effective potential is 
also shown (increasing from white to blue).}
\label{fig10}
\end{figure}
We have integrated a number of stellar orbits that all start at the
corotation radius $r=r_c$ with phases $\theta = 45^\circ,\,90^\circ,
... 225^\circ$. The results are shown in Fig. \ref{fig10} where we
also showed the effective potential
\begin{equation}
V_{\rm eff}(r,\theta) = V(r,\theta) - \frac{1}{2} \left(\Im(\omega)\right) ^2 r^2.
\end{equation}
All banana orbits evolve around the stable Lagrange point of the
system which is located at the local maximum of the effective
potential at a phase angle $\theta_L$. In the corotating reference
frame, a star at position ($r_c,\theta_L$) will remain there forever.
Orbits started at corotation but with different phase angles will
trace out a banana-shaped curve. The further away from the equilibrium
point $(r_c,\theta_L)$ an orbit is started, the more elongated the
banana becomes. The orbit started at $(r_c,180^\circ+\theta_L)$ is
circular in the corotating reference frame. The origin of the
one-armed spiral now becomes more clear. A weak, rotating lopsided
perturbation will catch stars close to the ($r_c,\theta_L$) Lagrange
point on banana orbits. These then spend more time near the stable
Lagrange point, creating an over-density there, and less time near the
diametrically opposite point, creating an under-density there. Thus,
the banana orbits help strengthen the lopsided mode, causing more
stars to be trapped in banana orbits, and so on.

\section{conclusions} \label{conclusions}

Using a toy dynamical model, we have investigated the properties and
causes of dynamical instabilities that may cause lopsidedness in disc
galaxies. 

We found the well-known {\em counter-rotation instability} to be the
dominant mode in disc galaxy models with strong counter-rotation and
small radial anisotropy. It is much stronger than any $m=2$ mode we
found in these models. The strength of this mode diminishes in models
with less counter-rotation and it eventually disappears in fully
rotating models. These, however, develop a different type of lopsided
mode, that becomes stronger as rotation increases, although it is
always much weaker than the $m=2$ spiral-arm mode. This instability
bears strong resemblance to the {\em eccentricity instability} that is
known to occur in gaseous and stellar near-Keplerian discs orbiting a
central massive object. We provide the first theoretical evidence,
based on a thorough mode-analysis of a suite of dynamical models for
disc galaxies embedded in a dark halo, that the eccentricity
instability can also occur in the fully pro-grade stellar discs of
spiral galaxies.

By integrating the orbits of an ensemble of stars in the perturbed
potential of the two extreme cases of full counter-rotation on the one
hand and full prograde rotation on the other hand, we investigated the
physics underlying the counter-rotation and eccentricity
instability. The counter-rotation instability grows by changing
near-circular orbits into aligned loop orbits that help maintain a
lopsided structure. If the non-linear regime, radial orbits are
changed into butterfly orbits. In the case of the eccentricity mode,
both radial and tangential orbits become aligned loops in a corotating
reference frame. In the non-linear regime, orbits near corotation are
trapped into resonance and describe banana-shaped figures in a
corotating frame. They help support the characteristic one-armed
spiral.

 In terms of density waves, the counter-rotating $m=1$ mode is
most likely due to a purely growing Jeans-type instability. An
approximative analytical criterion for local stability, akin to
Toomre's stability criterion for axisymmetric waves, can be employed
to estimate the region in which purely growing one-armed waves may
develop. This estimate roughly coincides with the region in which the
non-rotating lopsided mode is observed to reside. The rotating $m=1$
mode, on the other hand, grows as a result of the swing amplifier
working inside the resonance cavity that extends from the disc center
out to the radius where non-rotating rotating waves are stabilized by
the model's outwardly rising $Q$-profile. Rotating waves are confined
to even smaller radii so the non-rotating waves effectively set the
outer boundary of the resonance cavity.

Many disc galaxies show a noticeable $m=1$ perturbation besides the
dominant $m=2$ spiral-arm mode but only very few of them show any
counter-rotation. The rotating lopsided mode we identified in fully
prograde disc models therefore forms an attractive explanation for
this observed phenomenon.

\section*{Acknowledgments}

We thank the anonymous referee for his/her remarks that very much
improved the contents and presentation of the paper.

\label{lastpage}
\end{document}